\documentstyle[twocolumn,prb,aps]{revtex}
\begin{document}
\draft
\title{Magnetic Phase Diagram
 of the Ferromagnetically Stacked Triangular Ising Antiferromagnet}
\author{M.L. Plumer}
\address{ Centre de Recherche en Physique du Solide et D\'epartement de
Physique}
\address{Universit\'e de Sherbrooke, Sherbrooke, Qu\'ebec, Canada J1K 2R1}
\author{A. Mailhot}
\address{ INRS-EAU, 2800 rue Einstein, C.P. 7500,
            Ste.-Foy, Qu\'ebec, Canada G1V 4C7}
\date{November 1994}
\maketitle
\begin{abstract}
Histogram Monte-Carlo simulation results are presented for the
magnetic-field -- temperature phase diagram of the Ising model on a stacked
triangular lattice with antiferromagnetic intraplane and ferromagnetic
interplane interactions.   Finite-size scaling results for this frustrated
system at
three points along the paramagnetic transition boundary are presented which
strongly suggest a line of triciritcal points at low field and a first-order
transition line at higher fields.  These results are compared with the
corresponding phase diagrams from
conventional mean-field theory as well as from the Monte Carlo mean-field
calculations of Netz and Berker [Phys. Rev. Lett. {\bf 66}, 377 (1991)].
\end{abstract}
\section{Introduction}
Despite being perhaps the simplest example of a geometrically frustrated
magnetic system, and having a long history of investigation,
\cite{wan,copper,diep} the Ising antiferromagnet
on triangular and hexagonal lattices continues to reveal a number of remarkable
features\cite{netz} as well as elude concensus regarding some of its critical
properties.\cite{berk,olle,plum1,bunk,plum2}
The additional competition among interactions introduced by a
magnetic field often gives rise to a wide variety of multicritical-point
phenomena in frustrated systems, with a concomitant change in critical
behavior.
Although mean-field theories
usually can capture at least the qualitative structures of the magnetic
phase diagrams as determined by more sophisticated methods,
as well as by experiments,\cite{diep} there are notable exceptions, at least in
the case of the $XY$ model.\cite{plum3,plum4}
In this work, results of extensive Monte Carlo simulations of magnetic-field
effects on the stacked triangular Ising model with antiferromagnetic intraplane
and ferromagnetic interplane interactions are presented.  This study was
motivated by the work of Netz and Berker\cite{nb} ($NB$)
who employed a ``Monte-Carlo
Mean-Field" theory to investigate the phase diagram.  Although this method
is essentially based on a mean-field approximation, it preserves the
``hard-spin"
condition that is believed to be essential for frustrated systems.
In view of its relative simplicity, and the continuing controversy regarding
the nature of phase transitions in frustrated antiferromagnets,\cite{diep,xy}
Monte-Carlo Mean-Field theory is a powerful technique which deserves further
testing.

The magnetic ordering process of the frustrated Ising antiferromagnet
even at zero field has not been completely
resolved.  As the temperature is lowered from the paramagnetic ($P$) phase,
a transition to a partially disordered period-three state (phase $A$) occurs
where one-third of the sublattices remains paramagnetic.  Modified
mean-field theories
predicts that this phase remains stable down to zero temperature.\cite{mekata}
A second peak in the specific
heat at lower temperature observed in Monte Carlo ($MC$)
results has been interpreted as being associated with a transition to a
ferrimagnetic state (phase $B$).\cite{berk}  Such a
transition also has been indicated in the results of $NB$.  Other
reports of $MC$ simulations,\cite{mc} however, suggest that this peak is not
associated with a phase transition but signals the onset of new types
of excitations associated with the evolution of phase $A$ into the zero
temperature Wannier state.\cite{olle}  Notably, however, a low-temperature
phase $B$ is stabilized by the addition of next-nearest neighbor interactions,
with an intermediate (between phases $A$ and $B$) ferrimagnetic phase $C$,
as determined by a variety of mean-field
approximations\cite{mekata,plum5},
and as recently observed experimentally.\cite{far}

Magnetic-field effects have been much less studied.  The results of applying
a modified mean-field theory to the triangular lattice is a phase diagram
which exhibits a small region of phase $A$ at low fields, a large region of
phase $B$ at higher field strengths, and with the N\'eel temperature being a
multicritical point where phases $A$, $B$, and $P$ coincide.\cite{kab}
These results should also apply to the {\it ferromagnetically} coupled
triangular layered system of interest here since mean-field theory is
independent of space dimensions. The results of $NB$ for the
hexagonal system show two transitions at zero field, phase $A$
stabilized in a {\it small}
region at lower field values and higher temperatures, phase $B$ occupying most
of the phase diagram, but with a possibly complex multicritical point occurring
at a {\it finite}
field along the paramagnetic phase boundary.  An important aspect
of these results is the observation that the $P-B$ phase boundary is a
first-order transition line due to the 3-state (Potts) symmetry of phase $B$.
The transition at the N\'eel temperature is believed to be of $XY$ universality
(due to frustration),\cite{berk,plum2} and thus, so should be the entire
field-induced transition line $A-P$.  $NB$ also claim that the $A-B$ transition
is continuous.  However, we have argued that these two states have the same
symmetry so that the transition must be first order.\cite{plum5}  Other work
on phase diagrams, of possibly indirect relevance, was based on
a phenomenological model for the case of antiferromagnetic interlayer
coupling.\cite{plum5}

We study here the Ising Hamiltonian on a simple hexagonal lattice
\begin{equation}
{\cal H}~=~J_{\|} \sum_{<ij>} S_i S_j
+ J_{\bot} \sum_{<kl>}  S_k  S_l  - H \sum_i S_i
\end{equation}
where  $J_\|=-1$ is the ferromagnetic
interplane interaction, $J_{\bot}=1$ indicates
the antiferromagnetic intraplane coupling which is frustrated for
the triangular geometry, $<i,j>$
and $<k,l>$ represent near-neighbor sums along the hexagonal $c$ axis
and in the basal plane, respectively, with an applied field $H$.
After a preliminary examination of the phase diagram based on
a simplified molecular-field calculation\cite{diep} in the next section,
the results of finite-size scaling of conventional and
histogram\cite{ferr} MC data are presented.
The utility of these approaches in determining phase diagrams of
frustrated systems has been previously demonstrated.\cite{diep,plum4}
In contrast with other mean-field models, our calculations yield
a phase diagram where only phase $B$ is present, even at zero field.
Although the transition at $H=0$ is continuous,
the $P-B$ transition is found to be first-order for all $H \not= 0$.
The N\'eel temperature thus represents a tricritical point within this
approximation.  Interpretation of the MC data is more difficult due
to the ambiguity in making a
distinction between phases $A$ and $B$ in the presence
of a field, as well as the relatively large effects of critical
fluctuations.  It is clear, however, that phase $A$ is the only state
which occurs at $H=0$ (in agreement with other MC results and in contrast
with $NB$).
It is also evident that {\it if} this state exists at finite fields,
it occupies only a very small region of the phase diagram, roughly
consistent with the results of $NB$.  At high fields, the transition
to the paramagnetic state is weakly first-order,
in agreement with molecular-field
theory and symmetry arguments of $NB$.  At lower fields, however,
the behavior appears close to tricritical for both of the points
on the paramagnetic boundary that were examined.  This is in
contrast with the prediction of $XY$ universality ($NB$) (as found\cite{plum2}
at $H=0$) and suggests the possibility that phase $A$ exists
only {\it at} the paramagnetic boundary for $H \not= 0$.


\section{Molecular-Field Landau Theory}

Analyses of magnetic phase diagrams based on phenomenological Landau-type free
energies for frustrated spin systems have proven to be quite successful
in reproducing the essential features of both MC and experimental
results.\cite{diep}  Usually, qualitative features can be captured
using the more limited approach of deriving such a free energy from
a molecular-field treatment of Hamiltonians such as that given by (1).
This method allows for the understanding of relative phase stability
from analytic expressions and makes symmetry arguments clear.  For the
present work, the free energy is expanded to sixth order in the spin density
\begin{equation}
s(r)~=~  2[X \cos({\bf Q \cdot r} + \phi) + Y]
\end{equation}
where $X$ is the amplitude of the modulated component,
$Y$ is a uniform component, and ${\bf Q} = 4 \pi /(3a) \hat a$ is the
period-three (in the basal plane) wave vector.
Results for the free energy are similar to those of the phenomenological model
given in Ref.\onlinecite{plum5}
for the case of antiferromagnetic interlayer coupling,
but with a few important differences.
In the phenomenological model, the transition temperatures $T_x$ and
$T_y$, as well as the fourth and sixth-order coefficients, $B$ and $C$,
were treated as adjustable parameters.  Within the molecular-field
derivation, these quantities are determined by
\begin{equation}
T_X = (-2J_\| + 3J_{\bot})/a, ~~~ T_Y = -(2J_\| + 6J_{\bot})/a,
\end{equation}
along with $B=bT$ and $C=cT$, where the parameters $a$, $b$, and $c$
are given in terms of the total angular momentum $j$.\cite{plum6}
For simplicity, we take here $j=1$ so that, e.g., $a=3$.  In addition,
there is a direct Zeeman coupling $-HY$, absent in antiferromagnetic
interlayer-coupling case of Ref.\onlinecite{plum5}.

The three phases $A$, $B$, and $C$ (labelled $1$, $3$, and $2$, respectively
in Ref.\onlinecite{plum5}) are conveniently characterised by
$Y$ and the phase angle $\phi$: In phase A $Y=0$, $\phi = (2n+1)\pi /6$;
in phase B $Y\not=0$, $\phi = n\pi /3$; in phase C $Y\not=0$, and
$\phi$ takes on intermediate values.  Stabilization of the 3-state Potts
phase $B$ is due to a term third-order in $X$, $\sim YX^3\cos(3\phi)$, and
notably involves the uniform component $Y$, even in the absence of an
applied field.\cite{plum4}

The resulting phase diagram is shown in Fig. 1.  Only phase $B$ occurs,
even at zero field.  The transition $B-P$ is continuous at $H=0$ and is
first order otherwise.  The N\'eel transition thus represents a tricritical
point.

We note that within a corresponding phenomenological model, many
different types of phase diagrams can be constructed, including
those which feature all three types of ordered phases at zero field.
For the purposes of comparison with MC simulation results, only
the molecular-field calculation is given here.

\section{Phase Diagram from Monte Carlo Simulations}

Conventional Metropolis MC simulations were performed
on the Hamiltonian (1) in an effort
to determine the main features of the phase diagram.
Periodic boundary conditions were used on $L \times L \times L$ lattices
with $L$=18-30.
In some cases, a random initial spin configuration was used and
thermodynamic averages were estimated after discarding
the first $5-10 \times 10^4$ MC steps for thermalization.  In most cases,
however, the final well thermalized configuration of a previous run was used
for the initial spin directions.  Averages were then made using
$5-10 \times 10^5$ MC steps.  The paramagnetic phase boundary was
estimated by the location of extrema in a variety of thermodynamic functions,
discussed in the next section.  No such anomalies were observed at
the $A-B$ boundary line, which was crudely estimated at finite field
by the calculated value of the phase angle $\phi$.\cite{lee}
At $H=0$, a well-defined
signature of phase $A$ is the absence of the uniform component $Y$.
This quantity was found to extrapolate to zero at large $L$ for all
temperatures examined ($T > 0.4$) and no signature of a transition
($T \sim 0.9$)
to a low-T phase $B$, as found in Ref.\onlinecite{berk}, was observed.

The phase angle $\phi$ is not in general well defined since a multitude
of values are energetically equivalent, e.g., $\phi = \pi /6, \pi /2, ...$
in phase $A$.  $MC$ results at $H=0$ indicated that an average over these
values $(\langle \phi \rangle = \pi / 4)$ was produced,
suggesting a small energy barrier
between these states.  At finite field, however, a transition {\it region}
(dependent on $L$ and the number of $MC$ steps used) from this
averaged state to one with stabilized values of $\phi = n\pi /3$ was
observed.  We tentatively associate this observation with a possible
$A-B$ transition.  Energy barriers between the degenerate states of
phase $B$ are likely higher due to the presence of the cubic Potts term
discussed
above and it thus appears that a particular phase angle is selected for a
given $MC$ run.  Using $MC$ data at $L=21$,
a region of possible stability for phase $A$ determined in this way
is indicated in Fig. 2.  Note that the $Y$ is always non-zero in the presence
of a field due to the Zeeman interaction and thus is not a good indicator of
an $A-B$ transition.

Although there is general agreement (except at $H=0$) between the results
of Fig. 2 and those of $NB$, the existence of a finite region of
the ordered state $A$ in the phase diagram is suspect.  We base this comment
on the conspicuous absence of any anomalies in the thermodynamic
functions, as well as the finite-size results for critical exponents
presented in the next section: These do not support the predicted
$XY$ universality of the boundary line where the $A-P$ transition is
supposed to occur.

Additional structure found by $NB$ in the region of the multicritical
point where phases $A$, $B$, and $P$ meet is beyond the scope of the
present technique and thus could not be verified.

\section{Finite-Size Scaling}

The analysis of MC generated histograms used here
to determine finite-size scaling behavior is now well tested in standard
systems\cite{expon} as well as in frustrated Heisenberg,\cite{mail}
XY,\cite{xy} and Ising models.\cite{plum2}  In particular, new results for the
present model at $H=0$ yielded crtical-exponent estimates\cite{plum2}
$\beta = 0.341(4)$, $\gamma = 1.31(3)$, $\nu = 0.662(9)$, giving
$\alpha = 0.012(30)$ by the Rushbrooke equality.  These results
agree well with estimates from standard calculations on the unfrustrated $XY$
model, such as from renormalization group analysis:\cite{leg}
$\beta = 0.349(4)$, $\gamma = 1.315(7)$, $\nu = 0.671(5)$  (giving
$\alpha = -0.013(15)$).  Finite-size scaling at first-order transitions is also
now well understood\cite{first} and tested in frustrated systems.\cite{plum4}
In addition, tricritical behavior has been observed in exponent values
for the frustrated $XY$ model.\cite{xy}  In an effort to determine the
critical behavior of the paramagnetic phase boundary of Fig. 2, finite-size
scaling at three points (H=0.5, 1.0, and 3.0 as indicated on Fig. 2) was
performed using the same quality of $MC$ data as in Ref.\onlinecite{plum2}.

Scaling (of the form $\sim L^x$)
was performed at the critical temperature, $T_N$, estimated by the
order-parameter cumulant ($U_m$) crossing method, as used previously by us for
frustrated spin systems.\cite{plum2,xy,mail}
A variety of thermodynamic functions were examined, including the
specific heat $C$, order parameter $M$, susceptibility
$\chi \sim \langle M^2 \rangle$,
energy cumulant $U(T)~=~1 - \frac13 \langle E^4 \rangle /
\langle E^2 \rangle ^2$, and
the logarithmic derivative of $M$, which is equivalent to
$V_1(T)~=~ \langle ME \rangle / \langle M \rangle  - \langle E \rangle $,
as well as the second logarithmic derivative of $M$,
$V_2(T)~=~ \langle M^2 E \rangle / \langle M^2 \rangle  - \langle E \rangle $.
$MC$ runs were made on lattices $L=12-30$, with averaging performed using
$5 \times 10^5$ MC steps for the smaller
lattices, and up to $2 \times 10^6$ MC steps for the larger lattices,
after discarding the initial $1-5 \times 10^5$ MC steps for thermalization.
Additional averaging was then made using the results from between
6 (smaller lattices) to 12 separate runs.  This yields a reasonable
$2.4 \times 10^7$ $MC$ steps used for averaging at the largest lattice
size.  Gross estimates of error
bars, given as an indication of the general quality of the data,
were made by taking the standard deviation of results from these runs.

\subsection{Scaling at H=0.5}

Results of the histogram analysis performed at $H=0.5$ and $T=2.99$,
(near the paramagnetic phase boundary)
are presented here.  Fig. 3 shows data at which $U_m(T)$ for
$L$ =12 and 15 cross corresponding results at
larger values of $L'$, giving an estimate of the
critical temperature $T_N=2.987(5)$.  (This value compares with the result
at zero field, $T_N=2.9298(10)$.\cite{plum2})  Finite-size scaling of other
thermodynamic functions was performed at $T_N$, as well as at $T=2.982$ and
$T=2.992$, thus providing a method with which to estimate errors associated
with critical exponents.\cite{gott}
The extrapolated value of the energy cumulant, $U^*=0.66664(3)$,
shown in Fig. 4a, is close to the result $2/3$ expected at a continuous
transition.  Scaling of the other thermodynamic functions
is shown in Figs. 4b-e.

Critical exponent-ratios associated with these functions were estimated
using all the data, as well as with results at $L=12$ and $L=15$
excluded.\cite{gott}
Except for the specific heat, only a small effect was observed in the
estimated values.  By this method, $\alpha / \nu$ was found to be
0.85(4), 0.79(7), and 0.68(11), respectively, where quoted errors
represent only the robustness of the fits.  In contrast, the same
excercise yields 0.465(5), 0.455(4), and 0.451(5) for
$\beta / \nu$.  Results given on the figures are from fits made
with only the $L=12$ data excluded.  (Our conclusions are not affected
by which set of data is used in the estimates.)  The value $1 / \nu =
1.947(5)$ was also determined from the data for $V_2$ (not shown).
Our final results
at $H=0.5$ are $\alpha = 0.40(13)$, $\beta = 0.23(5)$, $\gamma = 1.09(12)$, and
(taking the average of the two estimates for $\nu$)
$\nu = 0.51(3)$, where errors take account of uncertainties in $T_N$.  These
values are not close to the expected $XY$ universality of an $A-P$
transition (as found at $H=0$).  They are suggestive, however, of
tricritical behavior, where $\alpha = 1/2$, $\beta = 1/4$, $\gamma = 1$, and
$\nu = 1/2$,

\subsection{Scaling at H=1.0}

Corresponding results at $H=1.0$ (based on histograms generated at $T=3.07$
for $L=12-27$ and $T=3.068$ for $L=30$) are presented in Figs. 5 and 6.
The critical temperature was estimated in this case to be $T_N=3.067(2)$.
Although the cumulant-crossing data for $L$=15 appears to be more widely
scattered in Fig. 5 than in Fig. 3, note the difference in temperature
scales by approximately a factor of two.  The possibility that this
transition is close to being first order is evidenced by the extrapolated
result $U^*=0.66656(8)$, which is slightly less than $2/3$ even within
error.  (As with all $MC$ data, however, caution must be taken in assessing the
reliabilty of error estimates.)   Following the method outlined above,
critical exponents at this transition were estimated to be
$\alpha = 0.59(14)$, $\beta = 0.22(3)$, $\gamma = 0.97(8)$, and
$\nu = 0.46(2)$ (with the result $1 / \nu = 2.17(2)$ from $V_2$ included).
Again, these results are not those expected of $XY$
universality, but (with the possible exception of $\nu$) are consistent
with tricritical behavior.

\subsection{Scaling at H=3.0}

At the relatively high field value of $H=3$, a first-order transition
from the paramagnetic to 3-state Potts phase $B$ is expected.
Significant scatter in the cumulant-crossing data of Fig. 7 is observed
(from histograms generated at $T=3.14$ for $L=12-27$ and $T=3.137$ for $L=30$),
but again note the increased sensitivity of the temperature scale
(a factor of three greater than that of Fig. 5).  The transition temperature
is estimated to be $T_N=3.1365(10)$, with an error notably smaller
than at the other two field values.

Based on our previous experience with weak first-order transitions in
frustrated spin systems, some care must be taken in the analysis and
interpretation of the data in the present case.\cite{plum4}  In particular,
finite-size effects appear to be especially important so that only
results at the larger lattice sizes may display the true critical
behavior.\cite{mail2}  This is evident in the data shown for the
energy cumulant, specific heat, and susceptibility ($\chi$) of Figs. 8a-c
where volume-dependent scaling is observed only at larger $L$.  Corresponding
results for the susceptibility defined
by $\chi' \sim \langle M^2 \rangle - \langle M \rangle ^2$ (considered to
be less useful at a continuous transition than $\chi$ as defined
above\cite{expon}), as well as $V_1$ and $V_2$, however, display volume
dependence even at the smaller lattices, as seen in Figs. 8d-e.
These behaviors, along with the extrapolated value $U^*=0.66484(15)$
(considerably less than $2/3$), are convincing indicators that the
transition is indeed first order and involves the $B$ phase.

\section{Conclusions}

Although general features of the phase diagram associated with the
ferromagnetically stacked triangular Ising antiferromagnet determined here by
extensive $MC$ simulations are in agreement with the Mean-Field Monte Carlo
results of Netz and Berker, some significant differences are evident.
In addition to finding only the partially ordered phase $A$ stable at
zero field, the results suggest that this phase is destabilized by
any finite field.  This suggestion is based primarily on finite-size
scaling results at the supposed $A-P$ transition where the expected $XY$
universality, found at $H=0$, is not observed in the critical exponents.
The results are, however, possibly consistent with this boundary in the phase
diagram being a line of tricritical points at which phases $A$, $B$ and $P$ are
degenerate.  We consider the possibility of non-universal critical behavior
(as a function of field) unlikely as this phenomenon, as far as we know,
has not previously been observed in a three dimensional system.

We find good evidence from finite-size scaling that the higher-field
transition to the 3-state Potts phase $B$ is indeed first order.  This
result in fully consistent with $NB$ and is also observed in the corresponding
$XY$ model.\cite{plum4}  Somewhat surprising was the degree to which
finite-size effects are important at this transition.  For this reason,
our original
intention of studying this model with the addition of site disorder in
an effort to observe a continuous 3-state Potts transition,\cite{berker}
has been abandonded in favor of studying the same effect in the corresponding
$XY$ model.\cite{plum4}  It is hoped that the present work will serve to
stimulate further theoretical effort as well as inspire a search for
suitable experimentally testable realizations.

\acknowledgements
We thank A.N. Berker for useful discussions.
This work was supported by NSERC of Canada and FCAR du Qu\'ebec.
%

\begin{figure}
\caption{Phase diagram from molecular-field theory.  Regions labelled
$P$ and $B$ represent paramagnetic and 3-state Potts phases, respectively.
Broken line is a first-order phase boundary.}
\label{fig1}
\end{figure}

\begin{figure}
\caption{Possible phase diagram from Monte Carlo simulations, labelled
as in Fig. 1, with the additional phase $A$ representing the partially
disordered state.  Question mark indicates our uncertainty regarding the
extent to which phase $A$ exists in the phase diagram.  Circles at H=0.5,
1.0, and 3.0 represent field values at which finite-size scaling was performed.
Squares denote possible locations of multicritical points.
Broken and solid lines represent first-order and continuous phase boundaries,
respectively.}
\label{fig2}
\end{figure}

\begin{figure}
\caption{Results of applying the cumulant-crossing method to estimate the
critical temperature at $H=0.5$, where b=L'/L.}
\label{fig3}
\end{figure}

\begin{figure}
\caption{Finite-size scaling at the paramagnetic transition for
$H=0.5$ at $T_N = 2.987$ where fitting is performed with the smallest
lattice (L=12) omitted.  Errors represent the robustness of the fit,
except in the case of the energy cumulant (a), where it is determined by the
uncertainty in $T_N$.  The specific heat, order parameter, susceptibility, and
logarithmic derivative of the order parameter are shown in (b)-(e),
respectively.}
\label{fig4}
\end{figure}

\begin{figure}
\caption{Results of applying the cumulant-crossing method to estimate the
critical temperature at $H=1.0$.}
\label{fig5}
\end{figure}

\begin{figure}
\caption{Finite-size scaling at the paramagnetic transition for
$H=1.0$ at $T_N = 3.067$, as in Fig. 4.}
\label{fig6}
\end{figure}

\begin{figure}
\caption{Results of applying the cumulant-crossing method to estimate the
critical temperature at $H=3.0$.}
\label{fig7}
\end{figure}

\begin{figure}
\caption{Finite-size scaling at the paramagnetic transition for
$H=3.0$ at $T_N = 3.1365$ with the assumption of volume dependence.}
\label{fig8}
\end{figure}

\end{document}